# Self-assembled Bismuth Selenide (Bi$_2$Se$_3$) quantum dots grown by molecular beam epitaxy

**Marcel S. Claro**[1‡], **Abhinandan Gangopadhyay**[2], **David J. Smith**[3], **Maria C. Tamargo**[1*]

[1]Department of Chemistry, The City College of New York, New York, NY 10031, USA and Ph.D. Program in Chemistry, The Graduate Center of the City University of New York, New York, NY 10016, USA

[2] School for Engineering of Matter, Transport and Energy, Arizona State University, Tempe, AZ 85287, USA

[3]Department of Physics, Arizona State University, Tempe, AZ 85287, USA

*mtamargo@ccny.cuny.edu

[‡]Current Address: INL – International Iberian Nanotechnology Laboratory, 4715-330 Braga, Portugal

## Abstract

We report the growth of self-assembled Bi$_2$Se$_3$ quantum dots (QDs) by molecular beam epitaxy on GaAs substrates using the droplet epitaxy technique. The QD formation occurs after anneal of Bismuth droplets under Selenium flux. Characterization by atomic force microscopy, scanning electron microscopy, X-ray diffraction, high-resolution transmission electron microscopy and X-ray reflectance spectroscopy is presented. The quantum dots are crystalline, with hexagonal shape, and have average dimensions of 12-nm height (12 quintuple layers) and 46-nm width, and a density of 8.5x10$^9$ cm$^{-2}$. This droplet growth technique provides a means to produce topological insulator QDs in a reproducible and controllable way, providing convenient access to a promising quantum material with singular spin properties.

## Introduction

The electronic confinement in all three dimensions of semiconductor quantum dots (QDs) leads to unique quantum properties. A discrete energy spectrum is produced, and the confinement impacts how the electrons interact with each other and to external influences, such as electric and magnetic fields. These quantum effects can be tuned by changes in the sizes of the dots or the strength of the confining potential. Such QDs are used in many diverse applications, spanning from devices such as lasers[1], solar cells[2] and photodetector[3], to the study of new physical phenomena, such as single photon interactions[4] and spin manipulation[5].

In the well-known common semiconductors, QDs are typically created using heterostructures defined by lithography or by self-assembled crystal growth. QD growths by molecular beam epitaxy (MBE) in Stranski-Krastanov mode[6] and by droplet epitaxy[7,8] are the most commonly used techniques for Si-Ge, and III-V semiconductors. QDs could also be formed by the electrostatic confinement of 2D electron gases[9]. However, in materials where the electronic states are protected by time-reversal symmetry, i.e., Graphene[10], and the class of materials known as Topological Insulators (TI), the electrostatic potential cannot confine or scatter electrons as usual, a property known as the Klein paradox[11]. Thus, quantum confinement in these materials can normally only be achieved by the formation of 3D heterostructures.

Three-dimensional TIs, such as $Bi_2Se_3$ and related materials, are insulators in the bulk form, usually with a narrow band gap. However, they have surface states with spins that are locked with momentum and protected by time-reversal symmetry[12]. These materials exhibit a tetradymite crystal structure with Se-Bi-Se-Bi-Se units, commonly referred to as quintuple layers, that are bonded together by van der Waals forces[13]. The applications of devices based on TIs include quantum computing, dissipation-less electronics, spintronics, enhanced thermoelectric effects and high performance flexible photonic devices[12]. Some properties of these materials, especially those related to spintronic and quantum computing, can be enhanced by QD confinement[14,15]. However, to our knowledge, very few experiments have been done in this direction. The lack of some means to fabricate mesoscopic TIs in a reproducible and controlled way is identified as a major obstacle. Among the many TIs, $Bi_2Se_3$ is particularly interesting since its band gap is larger than those of most other Tis, and the experimentally verified Dirac cone is in the gamma point[16].

$Bi_2Se_3$ has been grown successfully by MBE on different substrates[17], and a lithographically defined QD with quantum confinement was previously demonstrated[18]. Nonetheless, the MBE growth of $Bi_2Se_3$ occurs by Van der Waals epitaxy[19] since it is a layered material, and strain cannot be used to induce QD formation by the Stranski-Krastanov mode as in other materials.

In this work we demonstrate a viable method to create self-assembled quantum dots of $Bi_2Se_3$ by molecular beam epitaxy based on the droplet epitaxy technique.

## Results and Discussion

The samples were grown in a dual-chamber Riber 2300P system equipped with *in situ* reflection high-energy electron diffraction (RHEED). The GaAs semi-insulating (001) substrates were prepared by oxide desorption in the first chamber, followed by the growth of 200 nm of GaAs at 575 °C by conventional MBE. The substrates were cooled under an overpressure of Arsenic and a flat c(4x4) surface reconstruction, typical of As-terminated GaAs, was observed by RHEED at the end of this process. The substrates were then moved under ultra-high vacuum (UHV) to the second chamber for growth of the Bi droplets and the QDs. The background pressure of the second chamber where the droplet and QD growth occurred was $1–4 \times 10^{-10}$ Torr during growth. High-purity 6N bismuth (Bi) and selenium (Se) were provided by a RIBER double-zone cell for Bi and a Riber VCOR cracker cell for Se. Fluxes were measured by an ion gauge placed in the path of the fluxes. The estimated flux ratio was 10:1 Se to Bi. The temperature used for the Bi source was 730 °C and the beam equivalent flux (BEP) was $2 \times 10^{-8}$ Torr. These conditions led to an average growth rate of 32 nm/h for $Bi_2Se_3$ by van der Waals epitaxy.

The growth of the bismuth droplets was initiated by opening the Bi shutter when the prepared substrate temperature reached 250 °C, and the flat c(4x4) surface reconstruction was still visible. During the first few minutes, a circular ring started to appear in the RHEED pattern and the streaky pattern from GaAs started to fade, indicating the formation of amorphous material on the surface. The RHEED pattern became totally diffuse after 15 min with barely visible spots that appeared in some orientations. For the droplet sample, the Bi shutter was closed at that time and the sample was cooled down and extracted for analysis. For QD formation, after the Bi shutter closure, the Se shutter was opened and the sample was exposed to Se flux at the same temperature of 250 °C. The RHEED pattern changed from the diffuse pattern to a "spotty" (1x1) reconstruction pattern. After approx. 2 mins, the RHEED pattern was totally transformed into the pattern shown in Fig. 1. The droplets were exposed to the Se flux for 40 min in total to ensure the complete formation of the $Bi_2Se_3$ and to minimize Se vacancies. However, since the RHEED was stable after the first few minutes of growth, it should be possible to reduce this time considerably in future studies. The inset in Figure 1 illustrates the steps that occur during the QD growth.

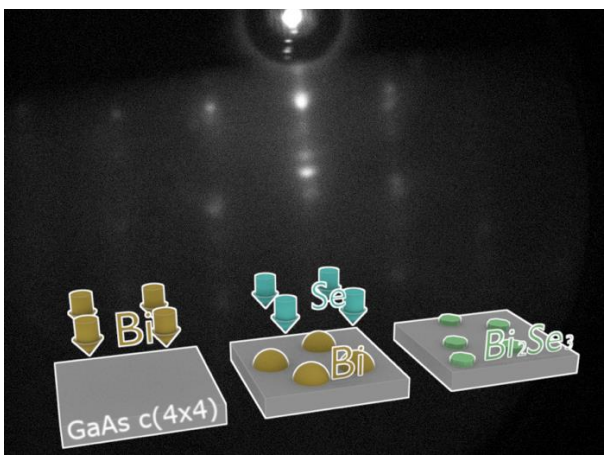

Figure 1: Spotty (1x1) RHEED pattern observed after formation of $Bi_2Se_3$ QDs. The inset illustrates the steps involved in the growth sequence.

Figure 2a) shows an atomic force microscope (AFM) image of the Bi droplets. They have semi-spherical shape, with average height of 29 nm and density of $4\times10^9$ $cm^{-2}$. Despite differences in the growth conditions, these results are comparable with Bi droplet growth reported in the literature[20].

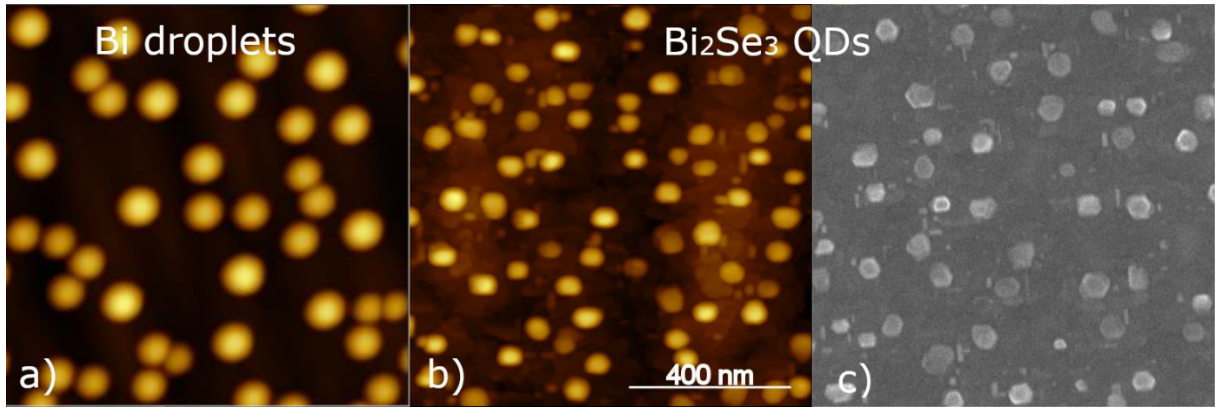

Figure 2: AFM images: a) bismuth droplets ; b) Bi$_2$Se$_3$ QDs. c) SEM image of Bi$_2$Se$_3$ QDs. All pictures have the same scale.

From the AFM image, (Fig. 2b), the QDs are measured to have an average height of 12 nm (12 quintuple-layers) and an average diameter of 46 nm, and the QD density is ~8.5x10$^9$ cm$^{-2}$. The size achieved is sufficient to preserve the characteristics of 3D TIs while, at the same time, some confinement is expected according to theoretical predictions[14] and previous measurements[18]. The transition to TI in thin films occurs at a thickness greater than about 6 quintuple layers[21]. In comparison to the radius, height, density, and finally the total volume of the Bi droplets, the QDs are smaller, shorter and more dense. Moreover, the QD density is approximately twice the droplet density. Several factors could explain this difference, such as the presence of some Bi desorption during the annealing process, ripening of droplets during QD formation, and changes in the Bi droplet size due to surface migration of Bi during cooling. These effects were observed in self-organized growth of nanostructures in other materials[22]. Due to convolution of the probe tip and the QDs, the QD shape is blurred in AFM images. However, from SEM images, as shown in Fig. 2c, the facets and the hexagonal shapes of the QDs are still preserved.

To establish the QD stoichiometry, the composition at the surface was measured using energy dispersive x-ray (EDX) analysis coupled to an SEM. Despite strong substrate signals from Ga and As, values of 2.3% of Bi and 3.9% of Se were found. The Se peak overlaps with the As peak causing small deviations in the Se content. Taking this into account, the ratio of Bi to Se from these measurements indicates the correct approximate stoichiometry for $Bi_2Se_3$.

High-resolution X-ray diffraction (HR-XRD) and X-ray reflection (XRR) experiments were also performed on the QDs. The results, after the alignment in the substrate (004) peak, are presented in Fig. 3a. The most intense $Bi_2Se_3$ peaks: (0003), (0006) and (00015) can be identified in the 2θ-ω scan. These demonstrate the presence of crystalline $Bi_2Se_3$ with the (0001) plane aligned to the (001) plane of the substrate. The noisiness and broadness of the peaks are not surprising due to the small amount of $Bi_2Se_3$ present on the surface and the inherent roughness of the QD layer. The roughness is also confirmed by the rapid damping observed in the measured XRR curve (Fig. 3a inset).

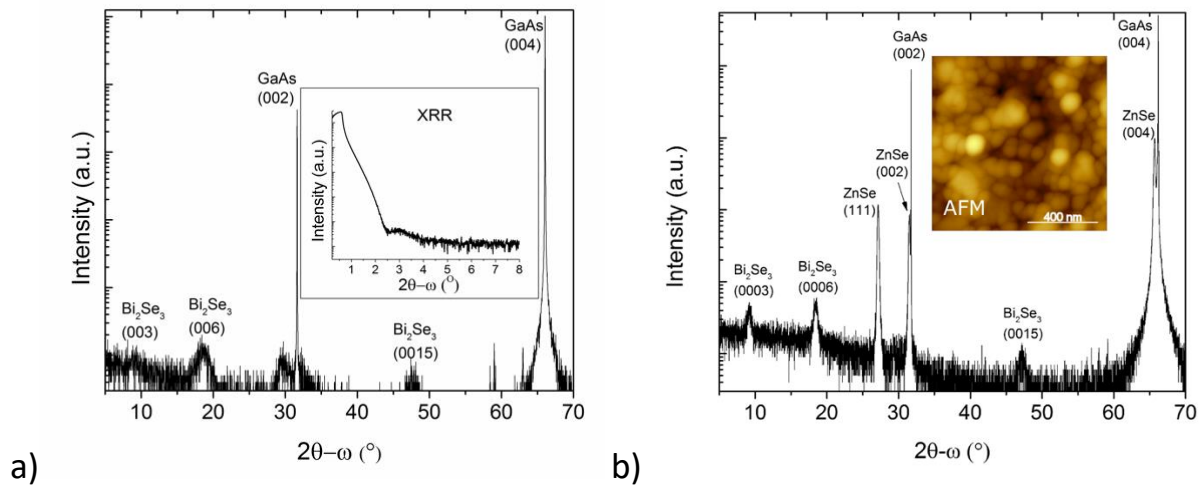

Figure 3: a) High-resolution X-Ray diffraction of $Bi_2Se_3$ QDs. X-ray reflection (XRR) is shown in the inset. b) High-resolution X-Ray diffraction pattern of $Bi_2Se_3$ QDs capped with 100 nm of ZnSe. The inset shows 1 μm x 1 μm AFM image of the surface.

Since the GaAs substrate surface remains exposed, any epitaxial material compatible with GaAs could be grown to cap the QDs. The capping layer will primarily protect the $Bi_2Se_3$ QDs from exfoliation and oxidation, and it could also serve as a barrier for carrier confinement. Unfortunately, $Bi_2Se_3$ will desorb from the surface when the sample temperature is raised above 300 °C. Thus, further growth of GaAs could not be used for capping purposes. Zinc Selenide, ZnSe, a wide-gap semiconductor ($E_g$ = 2.7 eV), has a very similar lattice parameter as GaAs and it can be grown as a strained layer on GaAs[23]. ZnSe has also been previously grown over $Bi_2Se_3$ layers[24]. The ideal temperature for ZnSe epitaxial growth is close to 250 °C, which is the temperature used for the QD growth.

Based on these considerations, another sample was grown where the QDs were capped with 100 nm of ZnSe. This growth took place in the same chamber, immediately after the QD growth. The zinc was provided by a RIBER Zn Knudsen effusion cell while the Se flux from the VCOR cell was maintained. The growth rate of ZnSe was 200 nm/h. The sample was exposed to Zn and Se fluxes for a period of time such that the overgrown layer thickness, assuming flat-layer growth, was ~100 nm. The HR-XRD pattern for the overgrown sample is shown in Fig. 3b. The $Bi_2Se_3$ peaks are still present as before, and narrow peaks corresponding to (002) and (004) of ZnSe can be observed adjacent to the corresponding GaAs peaks. The lattice mismatch relative to the GaAs is -0.57%, which means that the ZnSe layer should grow epitaxially and pseudomorphic to the GaAs substrate. From the AFM analysis, which is shown in the inset of Fig. 3b, it is notable that the surface is still rough even after growth of the capping layer (RMS smoothness ~ 7.4 nm) due to the initial roughness originating from the QDs, as well as ZnSe grains and defects. Furthermore, another peak is also visible in the HR-XRD pattern. This peak is attributed to the growth of grains of wurtzite ZnSe or of misoriented ZnSe on top of the $Bi_2Se_3$ QDs by van der Waals epitaxy[25]. Chen[25] and Hernandez-Mainet[24] *et al*, described the preferential growth of the wurtzite (hexagonal) phase when ZnCdSe was grown on $Bi_2Se_3$. However, we were unable to find the same described peaks or symmetry in the HR-XRD pattern. Instead, the peaks and symmetry observed indicate the presence of (111)-oriented ZnSe grains.

Aberration-corrected electron microscopy was used to image cross sections of the ZnSe-capped $Bi_2Se_3$ QDs: examples are shown in Figs. 4 and 5.  It is clear from Fig. 4 that the interface between GaAs and ZnSe is abrupt and continuous with few defects. Bismuth atomic columns appear with high contrast in both high-angle annular-dark-field (HAADF) and large-angle bright-field (LABF) image modes. Thus, the layered crystal (quintuple layer) structure of the $Bi_2Se_3$ QDs is readily identified in Fig. 4. Moreover, one can also clearly see the columnar growth of misoriented ZnSe, with stacking faults in the capping layer originating from the QDs or grain boundaries. Figures 5a and 5d clearly show the Bi2Se3 layered structure at very high resolution. The atomic stacking visible in the ZnSe capping layer directly above the $Bi_2 Se_3$ QD in Fig, 5a) suggests that ZnSe in this region grows as a disordered mixture of (0001) wurtzite and (111) zincblende structures, as previously observed in ZnSe nanobelt growth[26]  Fast Fourier Transforms (FFTs) of the TEM micrograph from areas above the QDs (Fig. 5c), on the sides of the QDs (i.e., on bare GaAs) (Fig. 5b), are very similar, despite the rotation, supporting the presence of tilted and defective zincblende ZnSe rather than the disordered wurtzite phase that is visible above the QDs. Finally, it is interesting that this image also suggests the presence of some bismuth at the grain boundaries, and the QD height is slightly greater close to the QD edge than in the QD center.  This implies the occurrence of some Bi segregation during the capping process, similar to what has been reported for Indium in the InAs SK QDs[27].

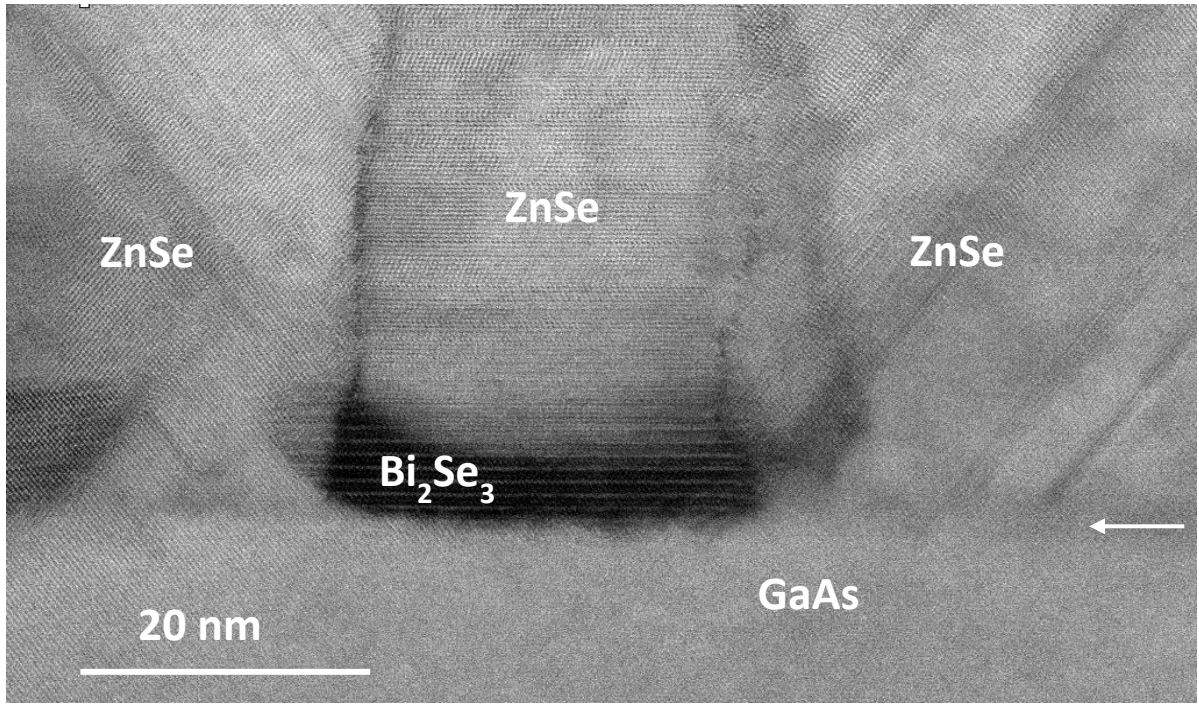

Figure 4: High-resolution scanning transmission electron micrograph recorded using large-angle bright-field imaging mode, showing $Bi_2Se_3$ QDs capped with 100nm of ZnSe. Arrow indicates top of GaAs substrate.

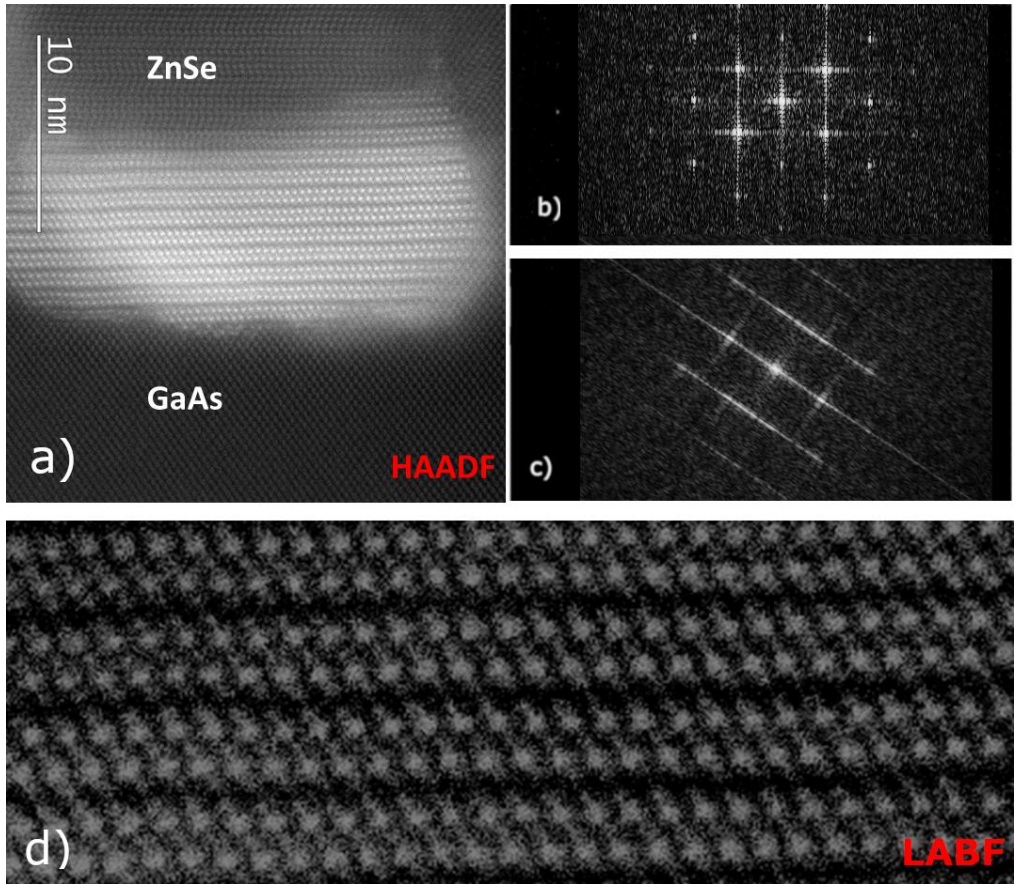

Figure 5: Aberration-corrected electron micrograph of $Bi_2Se_3$ QDs capped with 100nm of ZnSe: a) High-angle annular-dark-field (HAADF) image. b) FFT of GaAs region in (a). c) FFT of ZnSe region in (a) rotated by 45°. d) Detail of quintuple-layer stacking visible inside the $Bi_2Se_3$ QD. Inverted contrast large-angle bright-field (LABF) image.

## Conclusion

The droplet epitaxy technique has been successfully applied to the growth of quantum dots of $Bi_2Se_3$, which is a layered van der Waals material that is not amenable to more conventional strain-driven SK techniques of self-assembly via MBE. The $Bi_2Se_3$ QDs can be grown with high density and they have good crystal quality. Both uncapped and ZnSe-capped $Bi_2Se_3$ QDs were demonstrated. The size achieved is opportune, since electron confinement for a 3D Topological insulator would be anticipated. The QD formation process, as with other self-assembled techniques, is simple, reproducible and tunable, and should be ideal for the fabrication of large-area devices and experiments wherein the signal of several comparable quantum dots is required. The results offer the means to explore new physics and device applications possible with these unique low dimensional structures.

## Methods

The HR-XRD measurements were performed using a Bruker D8 Discover 3D diffractometer with a da Vinci configuration, 1-mm Slit/Collimator, and a Cu K$\alpha$1(1.5418 Å) source. The AFM images were captured with a Bruker Dimension FastScan AFM with a FastScan-A silicon probe and the SEM images with a FEI Helios Nanolab 660. The aberration-corrected imaging was carried out with a JEOL JEM-200F scanning transmission electron microscope operated at 200 keV. The probe convergence angle was set at 20 mrad, and the image collection angles were 0-22 mrad for large-angle bright-field imaging and 90-150 mrad for high-angle annular-dark-field imaging.

## References

1. Bimberg, D. & Pohl, U. W. Quantum dots: Promises and accomplishments. *Mater. Today* **14,** 388–397 (2011).
2. Sogabe, T., Shen, Q. & Yamaguchi, K. Recent progress on quantum dot solar cells: a review. *J. Photonics Energy* **6,** 040901 (2016).
3. Barve, A. V., Lee, S., Noh, S. & Krishna, S. Review of current progress in quantum dot infrared photodetectors. *Laser Photon. Rev.* **4,** 738–750 (2010).
4. Senellart, P., Solomon, G. & White, A. High-performance semiconductor quantum-dot single-photon sources. *Nat. Nanotechnol.* **12,** 1026–1039 (2017).
5. Yoneda, J. *et al.* A quantum-dot spin qubit with coherence limited by charge noise and fidelity higher than 99.9%. *Nat. Nanotechnol.* **13,** 102–106 (2018).
6. Joyce, B. A. B. A. & Vvedensky, D. D. D. Self-organized growth on GaAs surfaces. *Mater. Sci. Eng. R Reports* **46,** 127–176 (2004).
7. Reyes, K. *et al.* Unified model of droplet epitaxy for compound semiconductor nanostructures: Experiments and theory. *Phys. Rev. B - Condens. Matter Mater. Phys.* **87,** (2013).
8. Lee, J. *et al.* Various Quantum- and Nano-Structures by III-V Droplet Epitaxy on GaAs Substrates. *Nanoscale Res. Lett.* **5,** 308–14 (2009).
9. Ashoori, R. C. Electrons in artificial atoms. *Nature* **379,** 413–419 (1996).
10. Hewageegana, P. & Apalkov, V. Electron localization in graphene quantum dots. *Phys. Rev. B - Condens. Matter Mater. Phys.* **77,** 245426 (2008).
11. Katsnelson, M. I., Novoselov, K. S. & Geim, A. K. Chiral tunneling and the Klein paradox in graphene. (2006). doi:10.1038/nphys384

12. Hasan, M. Z. & Kane, C. L. Colloquium: Topological insulators. *Rev. Mod. Phys.* **82,** 3045–3067 (2010).
13. Ginley, T., Wang, Y. & Law, S. Topological Insulator Film Growth by Molecular Beam Epitaxy: A Review. *Crystals* **6,** 154 (2016).
14. Herath, T. M., Hewageegana, P. & Apalkov, V. A quantum dot in topological insulator nanofilm. *J. Phys. Condens. Matter* **26,** 115302 (2014).
15. Paudel, H. P. & Leuenberger, M. N. Three-dimensional topological insulator quantum dot for optically controlled quantum memory and quantum computing. *Phys. Rev. B* **88,** (2013).
16. Zhang, H. *et al.* Topological insulators in Bi2Se3, Bi2Te3 and Sb2Te3 with a single Dirac cone on the surface. *Nat. Phys.* **5,** 438–442 (2009).
17. Chen, Z. *et al.* Molecular Beam Epitaxial Growth and Properties of Bi2Se3 Topological Insulator Layers on Different Substrate Surfaces. *J. Electron. Mater.* **43,** 1–5 (2013).
18. Cho, S. *et al.* Topological insulator quantum dot with tunable barriers. *Nano Lett.* **12,** 469–472 (2012).
19. Walsh, L. A. & Hinkle, C. L. van der Waals epitaxy: 2D materials and topological insulators. *Applied Materials Today* **9,** 504–515 (2017).
20. Li, C. *et al.* Bismuth nano-droplets for group-V based molecular-beam droplet epitaxy. *Appl. Phys. Lett.* **99,** 1–5 (2011).
21. Xu, S. *et al.* van der Waals Epitaxial Growth of Atomically Thin Bi 2 Se 3 and Thickness-Dependent Topological Phase Transition. *Nano Lett.* **15,** 2645–2651 (2015).
22. Shchukin, V. A., Ledentsov, N. N. & Bimberg, D. *Epitaxy of Nanostructures*. *Springer Science* (Springer Berlin Heidelberg, 2003). doi:10.1007/978-3-662-07066-6
23. Kontos, A. G., Anastassakis, E., Chrysanthakopoulos, N., Calamiotou, M. & Pohl, U. W. Strain profiles in overcritical (001) ZnSe/GaAs heteroepitaxial layers. *J. Appl. Phys.* **86,** 412 (1999).
24. Hernandez-Mainet, L. C. *et al.* Two-dimensional X-ray diffraction characterization of (Zn,Cd,Mg)Se wurtzite layers grown on Bi2Se3. *J. Cryst. Growth* **433,** 122–127 (2016).
25. Chen, Z. *et al.* Molecular beam epitaxial growth and characterization of Bi 2 Se 3 /II-VI semiconductor heterostructures. *Cit. Appl. Phys. Lett. Appl. Phys. Lett. J. Vac. Sci. Technol. B Nanotechnol. Microelectron. Mater. Process. Meas. Phenom. J. Appl. Phys.* **1051,** 242105–21606 (2014).
26. Li, L. *et al.* Polarization-Induced Charge Distribution at Homogeneous Zincblende/Wurtzite Heterostructural Junctions in ZnSe Nanobelts. *Adv. Mater.* **24,** 1328–1332 (2012).
27. Keizer, J. G. *et al.* Kinetic Monte Carlo simulations and cross-sectional scanning tunneling microscopy as tools to investigate the heteroepitaxial capping of self-assembled quantum dots. *Phys. Rev. B* **85,** 155326 (2012).

# Acknowledgments

This work is supported by the National Science Foundation (NSF) CREST Center for Interface Design and Engineered Assembly of Low Dimensional systems (IDEALS), NSF grant number HRD-1547830 and by the NSF MRSEC for Precision Assembly of Superstratic and Superatomic Solids (PAS$^3$), NSF grant number DMR-1420634. The authors acknowledge the Nanofabrication and Imaging Facility of CUNY

## Author contribution statement

M.S.C conceived and executed the crystal growth. A.G. and D. J. S. performed the STEM experiments and analysis. M.S.C conducted the HR-XRD, XRR, AFM, SEM and EDX experiments. M.S.C., D. J. S. and M.C.T. wrote and reviewed the manuscript. All authors contributed to interpretation of the data and discussions.

**Competing interests:**
The authors declare no competing interests.

**Data availability statement:**

The datasets generated during and/or analyzed during the current study are available from the corresponding author on reasonable request.